%% file: tmbpaper.tex
\documentclass{iopart}
\usepackage{iopams,graphicx}
\include{definitions}

\usepackage[jphysicsB,abbr]{harvard}

\begin{document}
   \input{titelei.tex}

   \input{intro.tex}

   \input{theory.tex}

  \section{Numerical Results}\label{sec_results}
   \input{res_mav.tex}

   \input{res_0K.tex}
   \input{res_finT.tex}
   \input{conclusion.tex}

   \input{ack.tex}

  \section*{References}
   \bibliography{bec}

\end{document}

%% file: definitions.tex
\def\Tmbg{T_\mrm{mb}^\mrm{GHFB}}

\newcommand{\eps}{\varepsilon}

\newcommand{\eqref}[1]{(\ref{#1})}

\newcommand{\mrm}[1]{\ensuremath{\mathrm{#1}}}
\newcommand{\mb}[1]{\ensuremath{\mathbf{#1}}}

\newcommand{\intdl}[2]{\ensuremath{\int^{#2}_{#1}\!\mathrm{d}}}

\def\fB{\ensuremath{f_\mrm{B}}}

\def\T2b{\ensuremath{T_\mrm{2b}}}

\renewcommand{\inf}{\infty}
\def\bfr{{\bf r}}

\def\r{{\bf r}}

\def\q{{\bf q}}
\def\k{{\bf k}}

\def\Lop{\ensuremath{\hat{\mathcal{L}}}}                 

\def\dpsi{\delta\hat\psi(\bfr)}                 

\def\nc{\ensuremath{n_c(\bfr)}}
\def\nt{\ensuremath{\tilde{n}(\bfr)}}
\def\mt{\ensuremath{\widetilde{m}(\bfr)}}

\def\mav{\ensuremath{\widetilde{m}}}
\def\eps{\epsilon}

\def\kB{\ensuremath{k_\mathrm{B}}}

\def\wtrap{\ensuremath{\omega_\bot}}

\def\mavqm{\ensuremath{\mav_\mrm{qm}}}
\def\mavsc{\ensuremath{\mav'_\mrm{sc}}}
\def\mavr{\ensuremath{\mav'_\mrm{r}}}

\def\Be{Bose-Einstein}

\def\BdG{Bogoliubov-de Gennes}

\def\GPE{Gross-Pitaevski\u{\i} equation}

\def\ie{\mbox{i.\,e.\ }}

\newcommand{\ket}[1]{\ensuremath{|#1\rangle}}
\newcommand{\bra}[1]{\ensuremath{\langle #1|}}



%% file: titelei.tex

\title[Many-body T-matrix in GHFB]{Many-body T-matrix of a two-dimensional \Be\ condensate within the Hartree-Fock-Bogoliubov
 formalism}
\author{Christopher Gies\dag, M$\,$D Lee\ddag\ and D$\,$A$\:\!$W Hutchinson\dag}
\eads{\mailto{cluso@physics.otago.ac.nz}, \mailto{m.lee1@physics.ox.ac.uk},
\mailto{hutch@physics.otago.ac.nz}}

\address{\dag Department of Physics, University of Otago, P.O. Box 56, Dunedin, New Zealand}
\address{\ddag Clarendon Laboratory, Department of Physics, University of Oxford,
         Parks Road, Oxford OX1 3PU, United Kingdom}

\begin{abstract}
In a two-dimensional \Be\ condensate the reduction in dimensionality fundamentally influences
collisions between the atoms. In the crossover regime from three to two dimensions several
scattering parameters have been considered. However, finite temperature results are more
difficult to obtain. In this work we present the many-body T-matrix at finite temperatures
within a gapless Hartree-Fock-Bogoliubov approach and compare to zero and finite temperature
results obtained using different approaches. A semi-classical renormalization method is used
to remove the ultra-violet divergence of the anomalous average.
\end{abstract}

\pacs{03.65.Nk, 03.75.Hh, 03.75.Nt, 05.30.Jp}
\submitto{\JPB}

%% file: intro.tex
\section{Introduction}

When interaction processes within a Bose-condensed gas of atoms are considered,
collisions are often characterized by the two-body transition matrix
(T-matrix), evaluated in the limit of zero energy and collision momenta. In
three dimensions, this leads to the well-known expression
$g=4\pi\hbar^2a_\mrm{3D}/m$ for the coupling parameter, with $a_\mrm{3D}$ being
the three-dimensional, measured, \emph{s}-wave scattering length and $m$ the
mass of the atoms. The two-body T-matrix incorporates only binary collisions
and neglects effects that are due to the surrounding atoms. In three dimensions
these many-body effects have been shown to become important only in the regime
close to the critical temperature and are normally neglected
\cite{Hutchinson1998a}. In a two-dimensional system, however, scattering
processes are influenced by the reduction in dimensionality and the two-body
T-matrix differs from that in the three-dimensional case
\cite{Stoof_KT1993,Bijlsma1997a,Petrov2000a,Petrov2001long}. In the literature
various approaches have been invoked in order to obtain a valid coupling
parameter for a \Be\ condensate (BEC) in the two-dimensional regime. If
many-body effects are neglected, the interaction strength has been shown to
exhibit a logarithmic dependence on the collision energy
\cite{Schick1971,Fischer1988,Stoof1988,Kolomeisky1992,Stoof_KT1993,Bijlsma1997a,Kolomeisky2000a,Morgan2002b}
which goes to zero in the limit of zero collision energy. Therefore, in the
description of condensates, many-body effects dominate even in the first order
approximation, and so the determination of a many-body T-matrix which includes
these effects is of fundamental importance when studying two-dimensional BECs.

The Hartree-Fock-Bogoliubov (HFB) theory has proven successful in describing
finite temperature properties of dilute Bose gases 
\cite{Hutchinson1997a,Hutchinson2000a,Hutchinson2002_LaserPhys}. Many-body
effects on scattering processes are included in the theory through the
anomalous pair average. However, the full HFB theory is known to contain
various inconsistencies, such as a gap in the excitation spectrum and
ultra-violet divergences. To render the theory consistent, gapless extensions
have been developed \cite{Proukakis1998a,Hutchinson1998a,Hutchinson2000a}. In
this work, we use the gapless HFB theory, together with a semi-classical
renormalization procedure to remove the ultra-violet divergence of the
anomalous average, to study many-body effects on interactions. Many-body
effects due to the anomalous pair average can be neglected by taking the
simpler Popov approximation, which we have previously used to investigate the
failure of the semi-classical approximation in two dimensions \cite{Cluso1} and
to study the coherence properties of the two-dimensional BEC \cite{Cluso2}.
However, the problem of zero interaction strength in the two-body limit of the
coupling parameter persists and an appropriate effective contact interaction
strength has to be used.

Lee \etal have shown that, at zero temperature, the many-body T-matrix can be expressed
in terms of the two-body T-matrix, evaluated at a shifted off-shell energy
\cite{Lee2002a}. Recently, this approach has been extended to finite temperatures by
Rajagopal \etal for a homogeneous system \cite{Tosi2004a}. We compare results from both
these approaches at zero and finite temperatures to the results obtained from the gapless
HFB method, using local density approximations to incorporate the spatial dependence of
the condensate density in the trap.

The underlying theory is outlined in the following section. We briefly
summarize the gapless HFB method and how it incorporates many-body effects
before discussing in detail the renormalization of the anomalous average in
Section \ref{sec_renmtilde}. In Sections \ref{sec_tmb} and \ref{sec_finT_g2d}
we outline the meaning of the many-body T-matrix and how it can be expressed in
terms of two-body coupling parameters, depending on the regime in the
dimensional crossover from three to two dimensions, both at zero and finite
temperatures. In Section \ref{sec_results} we then present our numerical
results for the renormalized anomalous average and the many-body T-matrix.

%% file: theory.tex
 \section{Theory}

\subsection{Gapless Hartree-Fock-Bogoliubov formalism}

Within the HFB formalism the static properties of the BEC are described by a
coupled set of equations that require a self-consistent solution
\cite{Hutchinson2000a,Griffin1996b}. The order parameter $\Psi_0(\r)$ which
describes the condensed phase, obeys a generalized \GPE\ (GPE)
\begin{equation} \label{mf_gaplessGPE}
    \left( \hat h(\bfr)-\mu
   + g_\mrm{con}(\r)\,\nc + 2\, g_\mrm{exc}(\r)\, \nt\right) \Psi_0(\bfr) = 0~,\\
\end{equation}
while the thermal cloud is determined by the quasiparticle amplitudes,
$u_i,\,v_i$, and quasiparticle energies, $E_i$, which obey the coupled
\BdG\ (BdG) equations
\begin{eqnarray}\label{mf_gaplessBdG}
  \Lop\,u_i(\bfr) -\mathcal{M} \, v_i(\bfr) = E_i\, u_i(\bfr)\nonumber \\
  \Lop\,v_i(\bfr) -\mathcal{M} \, u_i(\bfr) = - E_i\, v_i(\bfr)~.
\end{eqnarray}
Here
\begin{equation}
 \label{mf_gaplessLop}  \Lop = \hat h(\bfr)-\mu +
         2\, \Big( g_\mrm{con}(\r)\,\nc + g_\mrm{exc}(\r)\, \nt \Big),
\end{equation}
and,
\begin{equation}
 \label{mf_gaplessM} \mathcal{M} =  g_\mrm{con}(\r)\,\nc~.
\end{equation}
$\hat h(\r)=-\hbar^2\Delta/2m + U_\mrm{trap}$ is the single particle
Hamiltonian, and $n_c(\r)$, $\nt$ and $n(\bfr)=n_c(\bfr)+\nt$ refer to the
condensate, non-condensate and total densities respectively. In
\eqref{mf_gaplessGPE} and \eqref{mf_gaplessBdG} we have written the HFB
equations in a generalized form which, as we point out, is not precisely the
full HFB theory as derived in \cite{Griffin1996b}. It has been shown that the
occurrence of the anomalous pair average $\mt = \langle \dpsi \dpsi \rangle$
leads to a gap in the excitation spectrum \cite{Griffin1996b}. This, however,
is forbidden by Goldstone's theorem \cite{Goldstone1961} or the more general
Hugenholtz-Pines theorem \cite{Hugenholtz1959a}. Several resolutions to this
problem exist. One can neglect the anomalous average completely; the so-called
Popov approximation. However, the anomalous average has been shown to introduce
many-body effects via the many-body T-matrix for scattering processes
\cite{Proukakis1998a,Morgan2000a} through
\begin{equation}\label{mf_VconTMB}
     T^\mrm{GHFB}_\mrm{mb}(\r)= g\,\left( 1+ \frac{\mt}{\Psi_0(\r)^2} \right)~.
\end{equation}
Thus, the Popov approximation neglects these many-body effects. The problems from which
the full HFB theory suffers stem from the inconsistent introduction of these effects.
Gapless extensions to the full HFB theory have been developed that do not neglect the
anomalous average, but rather render the theory consistent by making sure these effects
are accounted for equally for all classes of scattering processes. These theories have
been termed G1 and G2 and can be identified by the definition of the coupling parameters
in \eqref{mf_gaplessGPE} and \eqref{mf_gaplessBdG} \cite{Proukakis1998a,Hutchinson2000a},
\ie
\begin{eqnarray}
 g_\mrm{con}(\r) = \Tmbg(\r),\quad g_\mrm{exc}=g \qquad \qquad \quad &\mbox{for G1}\\
 g_\mrm{con}(\r) = \Tmbg(\r),\quad g_\mrm{exc}(\r)= \Tmbg(\r)&\mbox{for G2}~.
\end{eqnarray}
The G1 theory includes many-body effects only for collisions within the
condensate and has been shown to be justified by a perturbative second order
treatment of the full HFB theory \cite{Morgan2000a}. The G2 extension also
includes many-body effects for condensate--non-condensate interactions, and has
been applied successfully to explain the downward shift in the $m=2$
excitation  frequency for an anisotropically trapped three-dimensional BEC
\cite{Hutchinson1998a} at temperatures approaching the critical temperature.
Note that equation \eqref{mf_VconTMB} requires the knowledge of the coupling
parameter $g$ in the two-body limit. We discuss the determination of $g$ in
Section \ref{sec_g}.

Both the non-condensate density and the anomalous average are calculated by populating
the quasiparticle levels via the Bose distribution function:
\begin{eqnarray}
 \nt = \sum_i
    \fB(E_i)\,\Big(|u_i(\bfr)|^2+|v_i(\bfr)|^2 \Big)+|v_i(\bfr)|^2\label{mf_nth-uv}\\
 \mt = \sum_i
    \Big(2\fB(E_i)+1\Big)\,u_i(\bfr)v_i^*(\bfr)~.\label{mf_mt-uv}
\end{eqnarray}
The Bose distribution function with the inverse temperature $\beta = 1/k_BT$ is
given by $\fB(E_i) = [{z^{-1} e^{\beta E_i} -1}]^{-1}$ with the fugacity
$z^{-1} = 1+N_0^{-1}$.

While the summations in \eqref{mf_nth-uv} and \eqref{mf_mt-uv} are infinite, when calculating
numerical solutions to this system of equations, we are compelled to introduce a high energy
cut-off $\eps_\mrm{cut}$.  We use a semi-classical treatment \cite{Reidl1999a,Cluso2} to
calculate the contributions to $\tilde{n}$ and $\tilde{m}$ from above the cut-off, and also to
perform a necessary renormalization of the anomalous average, as discussed in the following
section.

\subsubsection{Renormalization of the anomalous average.}\label{sec_renmtilde}

To go beyond the Popov approximation, the many-body T-matrix must be calculated
from \eqref{mf_VconTMB}. In order to obtain a divergence free anomalous
average, the vacuum contributions already included in the measured value of the
\emph{s}-wave scattering length must be explicitly subtracted. To motivate the
renormalization procedure, we briefly discuss its origin through the relation
between the anomalous average and the many-body T-matrix. In the homogeneous
limit the off-diagonal matrix element $\mathcal{M}$ \eqref{mf_gaplessM}, which
is defined in terms of the bare interaction potential $V(\r-\r')$, obeys a
Lippmann-Schwinger equation and can be identified with the many-body
T-matrix. In the pseudo potential approximation \cite{HuangStatMech}, the
matrix elements of the bare interaction potential are replaced by the contact
interaction $\delta$-potential, which leads to ultra-violet divergences in the
gapless HFB theories.  In fact, the contact potential approximation is better
applied to the two-body T-matrix. Introducing the two-body T-matrix and taking
the contact potential approximations leads to the inclusion of an additional
term which, together with $\mav$, we identify as a renormalized anomalous
average $\mav^R$. Replacing the bare interaction potential by the two-body
T-matrix is consistent to all orders in the Lippmann-Schwinger equation which
defines the many-body T-matrix if it is accompanied by this renormalization
\cite{Proukakis1998a,Morgan2000a}.  The equations which give the gapless HFB
theories are unchanged, save for the replacement of the anomalous average by
$\mav^R$ everywhere that it appears.

We now describe how we obtain the divergence free anomalous average. To simplify the
formulation within this section, we omit the notation of spatial dependencies and
introduce the following notation for the different parts of the anomalous average
\begin{eqnarray}\nonumber
    \mav^R &= \sum_i^{E_i<\eps_\mrm{cut}} \big(2\fB(E_i)+1\big)\, u_i v_i \\
     &\quad +\intdl{\eps_\mrm{cut}}{\inf}E\; \left\{\mav'_\mrm{sc}(E) -
                                          \mav'_\mrm{r}(E) \right\}
            -\intdl{0}{\eps_\mrm{cut}}E\; \mav'_\mrm{r}(E)\nonumber \\
      & \equiv \mav_\mrm{qm}  + \mav_1 - \mav_2~. \label{nm_mav_notation}
\end{eqnarray}
Here, the first part $\mav_\mrm{qm}$ is calculated quantum mechanically by
explicitly evaluating the sum in \eqref{mf_mt-uv} up to the energy cutoff. The
second part, $\mav_1$, contains the contribution from the energy levels above
the cutoff $\eps_\mrm{cut}$, given by subtracting the renormalization term
$\mav'_\mrm{r}$ from the semiclassical anomalous average integrand
$\mav'_\mrm{sc}$. The third part, $\mav_2$, is the renormalization of \mavqm.
Both $\mav_1$ and $\mav_2$ are calculated within the semi-classical
approximation.

In analogy to the semi-classical expression for the non-condensate density in two
dimensions, see \cite{Cluso2}, the unrenormalized semi-classical expression for the
anomalous average integrand can be found to be
\begin{eqnarray}\nonumber
    \mav'_\mrm{sc}(E) &= - \frac{m}{2\pi\hbar^2}\,
     \big(2\,\fB(E)+1 \big) \, \frac{g n_c}{\sqrt{E^2+(g n_c)^2}} \\
     & \quad \times \Theta\left( E-\sqrt{\left(U_\mrm{trap}-\mu+2gn\right)^2 -
        \left(gn_c\right)^2}\, \right)~.\label{mf_mtilde_sc}
\end{eqnarray}
The integral is logarithmically divergent for the non-thermal part on the upper
boundary, but this divergence is removed by the renormalization. To calculate
the excessive contribution to the anomalous average, we consider the case
$n_c \rightarrow 0$. In this sense we keep only terms linear in the condensate
density. This arises from the assumption that we renormalize the vacuum
contributions to $\mav$ at zero temperature that have already been accounted
for in the measured scattering length in the two-body T-matrix. In the limit
that there is no condensate present, the quasiparticle energies are replaced by
the single particle energies. Since the quasiparticle energies are measured
relative to the condensate, we replace
\begin{equation}\label{nm_m:replaceE}
    E \quad \rightarrow \quad E + \mu~.
\end{equation}
Doing so in \eqref{mf_mtilde_sc} and regarding only the temperature independent
part leads to
\begin{equation}\label{nm_m:mt_ncsmall}
    \mavr(E) = - \frac{m}{2\pi\hbar^2} \:\frac{g n_c}{E +\mu}~.
\end{equation}
%

\paragraph{Above the Energy Cutoff.}
We can now calculate the semi-classical contribution to $\mav^R$ from above the energy
cutoff, \ie
\begin{equation}\label{nm_m:mt1}
    \mav_1 = \intdl{E_\mrm{min}}{\inf} E\; \big(\mavsc(E)-\mavr(E)\big)~.
\end{equation}
The lower limit of the integral is determined by the maximum of the Heaviside
function in \eqref{mf_mtilde_sc} and the cutoff energy, \ie
\begin{equation}\label{nm_m:ll}
    E_\mrm{min} = \max \bigg\{\eps_\mrm{cut},\sqrt{\left(U_\mrm{trap}-\mu+2gn\right)^2 -
        \left(gn_c\right)^2}\, \bigg\}~.
\end{equation}
Note that the shift of the energy in the denominator of \eqref{nm_m:mt_ncsmall}
prevents the integral from being singular at the upper limit.

Since the semi-classical part \eqref{mf_mtilde_sc} of the anomalous average is
divergent at the upper boundary, the renormalization \eqref{nm_m:mt_ncsmall}
must be subtracted under the same integral, so that we get for \eqref{nm_m:mt1}
\begin{equation}\label{nm_m:msc}
    \mav_1 = - \frac{m}{2\pi\hbar^2} \int_{E_\mrm{min}}^\inf\!\!\!\! \mrm{d}E\;
    gn_c\,\left(\frac{2\,\fB(E)+1}{
    \sqrt{E^2-(g n_c)^2} } \,
    - \frac{1}{E +\mu}~ \right) .
\end{equation}
%

\paragraph{Below the Energy Cutoff.}
Since the sum in \mavqm\ has a natural cutoff it does not diverge in the
calculation. However, it contains contributions from the zero-temperature limit
that must be renormalized by subtracting the integral of \mavr\ below the
energy cutoff. In analogy to \eqref{nm_m:mt_ncsmall}, this is given by
\begin{equation}\label{nm_m:mrenormlow}
    \mav_2 = - \frac{m}{2\pi\hbar^2} \int_{0}^{E_\mrm{min}}\!\!\!\! \mrm{d}E\;
    \frac{g n_c}{E +\mu}~.
\end{equation}
%


\subsection{Determination of the many-body T-matrix}\label{sec_tmb}

In three dimensions the effect of the surrounding medium on the scattering of two atoms
plays a role only at high temperatures \cite{Hutchinson1998a}. However, in the lower
dimensional case these effects must be accounted for even at zero temperature, leading to
the introduction of the many-body T-matrix \cite{Bijlsma1997a,Proukakis1998a,Lee2002a}.

The many-body T-matrix used here is defined by the Lippmann-Schwinger equation
\begin{eqnarray}
    \fl
    \bra{\k'}T_\mrm{mb}(E)\ket{\k} =& \bra{\k'} V(\r) \ket{\k}
    \nonumber \\ \label{cp:TMb=LippS}
     &+ \sum_\mb{q} \bra{\k'}V(\r)\ket{\mb{q}} \,
     \frac{1+n_\q+n_{-\q}}
     {E-(E_{\q} + E_{-\q})}
     \, \bra{\mb{q}} T_\mrm{mb}(E)\ket{\k}~.
\end{eqnarray}
This differs from the two-body T-matrix by the use of the quasiparticle energy spectrum
for the intermediate states in a collision, and by the Bose enhancement of scattering
into these states, given by the population terms. In the case of BEC, usually the zero
momentum and energy limit is considered.

In the previous section we have briefly mentioned that the many-body T-matrix can be
calculated by means of the anomalous average and the condensate wave function.  Lee \etal
\cite{Lee2002a} have shown that in two dimensions, at zero temperature, the many-body
T-matrix can be obtained from the known expression for the two-body T-matrix at an
off-shell shifted collision energy, \ie
\begin{equation}\label{tmb_t2b}
 T_\mrm{mb} = T_\mrm{2b}(-\mu)~.
\end{equation}
This result includes the effect of the quasiparticle energy spectrum of the intermediate
states in the collision and provides a convenient method of calculating the many-body
T-matrix. This should not be confused with the approach recently used by other authors
\cite{Khawaja2002a,Tosi2004a,Stoof_KT1993}, who have also used a shifted energy in the
two-body T-matrix, namely an energy of $-2\mu$. The latter result stems from the argument that
the excitation of a single condensate atom is associated with an energy of $-\mu$, so that for
a condensate--condensate interaction the energy of the collision is then $-2\mu$. Such an
argument includes only the mean-field energy of the initial and final states, and neglects the
other many-body effects on the collision which are included in the result in \eqref{tmb_t2b}.

The off-shell two-body T-matrix has been calculated for the genuinely two-dimensional
case \cite{Morgan2002b}, and using \eqref{tmb_t2b} we can obtain the many-body T-matrix.
Experimental realizations of a two-dimensional system are not genuinely two-dimensional
however, and different regimes in the dimensional crossover must be distinguished, which
we now discuss.

\subsubsection{Two-body coupling parameter.}\label{sec_t2b}

Interactions in a three-dimensional condensate are usually parameterized by the three
dimensional coupling parameter $g_\mrm{3D}=4\pi\hbar^2a_\mrm{3D}/m$, which is the
zero-temperature and zero-energy limit of the two-body T-matrix in three dimensions.

As the axial confinement of the trap is tightened, so that the dynamics of the
atoms in the axial direction are frozen out, the dimension of the condensate is
reduced from three to two dimensions. However, since scattering processes take
place on a very small length scale, they can still be considered to take place
in three spatial dimensions as long as the three dimensional scattering length
$a_\mrm{3D}$ is much smaller than the characteristic length of the trap,
$l_z=\sqrt{\hbar/m \omega_z}$, with $\omega_z$ the axial trapping frequency.
During this crossover, the system can be classified into three different regimes:
quasi-3D, quasi-2D and genuine 2D, where the latter, in which scattering itself
is purely two-dimensional, is as yet inaccessible to experiment.

In the quasi-3D regime where $l_z \gg a_\mrm{3D}$, the axial $z$-component of
the condensate wave function can be assumed to be Gaussian. Scattering is not
directly affected by the axial confinement, and the tightly confined part can
be factorized from the wave function, leading to a two-dimensional GPE with a
modified interaction strength \cite{Lee2002a}, \ie
\begin{equation}\label{nm_g_3Dprime}
    g_\mrm{q3D} = \sqrt{\frac{m \omega_z}{2\pi \hbar}} \; g_\mrm{3D}~.
\end{equation}

If the axial confinement is tightened further, collisions are influenced by the
reduced dimensionality and the scattering problem must be considered
explicitly. The quasi-2D regime, where $l_z \gtrsim a_\mrm{3D}$, has previously
been investigated by Petrov \etal \cite{Petrov2001long,Petrov2000a}. They found
that the coupling parameter depends logarithmically on the tightness of the
confinement and the collision energy. This regime is currently accessible to
experiments \cite{Goerlitz2001_2D} and forms the case we consider in the
remainder of this paper. In order to obtain the many-body T-matrix, we evaluate
the two-body coupling parameter at the negative energy $-\mu$ corresponding to
\eqref{tmb_t2b}. For the homogeneous gas this can be written as \cite{Lee2002a}
\begin{equation}\label{nm_g2D_TMB}
    g_\mrm{q2D} \equiv T_\mrm{mb}(E=0) =  \frac{4\pi\hbar^2}{m}\,
     \frac{1}{\ln\big(4 \hbar^2/\mu m a^2_\mrm{2D} \big)}~,
\end{equation}
where the two-dimensional scattering length is given in terms of the three
dimensional scattering length and the parameter $l_z/a_\mrm{3D}$ by
\begin{equation}\label{cp:a2D}
    a_\mrm{2D} = 4\,\sqrt{\frac{\pi}{B}} \:l_z\,e^{-\sqrt{\pi} \,
    \frac{\scriptstyle{l}_z}{\scriptstyle{a}_\mrm{3D}}} ~,
\end{equation}
with the constant $B\approx 0.915$. Note that \eqref{nm_g2D_TMB} is valid strictly only
at zero temperature. Because this form of the many-body T-matrix is valid in the quasi-2D
regime, we refer to \eqref{nm_g2D_TMB} with \eqref{cp:a2D} as the \emph{quasi-2D coupling
parameter}.

In the genuine two-dimensional limit, where $l_z \lesssim a_\mrm{3D}$, the scattering
length $a_\mrm{2D}$ in \eqref{nm_g2D_TMB} is a purely two dimensional quantity and cannot
be approximated by \eqref{cp:a2D}. Since this regime is currently not experimentally
accessible, we do not discuss details, but refer to \cite{Tosi2002a,Lee2002a}.

The expressions given above are strictly valid only for a gas which is homogeneous in the
two relevant dimensions. Provided that the trapping potential varies slowly on the length
scale on which interactions take place, we can use the zero-temperature relation $\mu =
gn_c(\r)$ to replace the chemical potential within a local density approximation. This
results in the spatially dependent coupling parameter in the quasi-2D regime
\begin{equation}\label{cp:g2d_spatial}
    g_\mrm{q2D}(\r) =
    \frac{4\,\pi \hbar^2}{m} \,\frac{1}{\ln\big(4 \hbar^2/n_c(\r)\,g_\mrm{q2D}(\r)\, m a_\mrm{2D}^2\big)}
\end{equation}
which can be crudely approximated by
\begin{equation}\label{cp:g2d_spatial_approx}
    g_\mrm{q2D}(\r) \approx
    - \frac{4\,\pi \hbar^2}{m} \,\frac{1}{\ln\big(n_c(\r)\,\pi a_\mrm{2D}^2\big)}~.
\end{equation}
This approximation overestimates \eqref{cp:g2d_spatial}, but we use it as a
starting value in our first iteration to solve \eqref{cp:g2d_spatial}
self-consistently.


\subsubsection{Coupling strength $g$ in gapless HFB.}\label{sec_g}

The interaction strength $g$ in \eqref{mf_VconTMB} is defined by the two-body
coupling parameter. In two dimensions this is given by \eqref{nm_g2D_TMB} in
the limit $\mu=E=0$ and hence vanishes, thus the determination of $g$ is not
straightforward. Naturally, since the many-body T-matrix from the gapless HFB
approach must agree with the quasi-2D coupling parameter from the previous
section, we can use this to determine the interaction strength $g$. Equations
\eqref{nm_g2D_TMB} and \eqref{cp:g2d_spatial} are valid at zero temperature,
and we solve the HFB equations self-consistently at zero temperature to
determine $g$ so that the gapless HFB many-body T-matrix agrees with the
quasi-2D coupling parameter \eqref{cp:g2d_spatial} at the trap centre.

\subsubsection{Extension to finite temperatures}\label{sec_finT_g2d}

The quasi-2D coupling parameter given in Section \ref{sec_t2b} was derived
using the zero temperature relationship of \eqref{tmb_t2b}.  As a consequence,
the effect of the population terms in \eqref{cp:TMb=LippS} have been neglected.
At finite temperatures these population terms will become significant and an
extension to the results of Section \ref{sec_t2b} is required.  Such an
extension presents a significant problem.  In the approach outlined in Section
\ref{sec_t2b} the many-body T-matrix was calculated first in a homogeneous 2D
zero-temperature system, before using the local density approximation to extend
the results to the trapped case.   At finite temperatures, however, a
condensate cannot exist in a 2D homogeneous system. As a consequence, the
calculation of the homogeneous finite temperature many-body T-matrix is plagued
by infra-red divergences.  In principle, the correct procedure would require
the solution of the many-body scattering problem in a trap, but this is
extremely difficult.  A potential extension which avoids the problem of
divergences was proposed in a recent publication by Rajagopal \etal
\cite{Tosi2004a}.  In their work the quasiparticle spectrum is approximated by
$E_i \approx E_i^{sp} +\mu$ (where $E_i^{sp}$ is the single-particle energy
spectrum), which is valid for the high energy states, but which neglects the
phonon part of the spectrum.  Using this approximation, the finite temperature
result for the coupling parameter of Rajagopal \etal is given by
\begin{equation}\label{cp:g2d_finT_tosi}
    g_\mrm{q2D}(T) =
     \frac{4\,\pi \hbar^2}{m} \,\frac{1}{\ln\left(\frac{4\hbar^2}{\mu
     a_\mrm{2D}^2}\right)
     -2\sum_{s=1}^{\inf}\mrm{Ei}\, \big(\frac{- s\mu}{\kB T}\big)}~.
\end{equation}
where $\mrm{Ei}(x)$ is the exponential integral function.  In order to extend
\eqref{cp:g2d_finT_tosi} to trapped condensates, we replace the chemical
potential within a local density approximation which includes the thermal
atoms, \ie
\begin{equation}\label{finT_lda}
\mu \rightarrow g_\mrm{q2D}[n_c(r)+2\tilde n(r)]~.
\end{equation}
In the following section we will compare this result to the finite-temperature
numerical results of HFB theory.

%% file: res_mav.tex
We now present the results of our numerical calculation. We use the G2 formalism and the
numerical methods are outlined in \cite{Hutchinson2000a,Cluso2}. All quantities are shown
in trap units, \ie lengths are scaled by the radial harmonic oscillator length
$l_\bot=\sqrt{\hbar/m\wtrap}$ and energies by $E_0=\hbar\wtrap/2$. We consider a sample
of 2000 sodium atoms. The critical temperature of the non-interacting gas is
approximately $33\,$nK.

\subsection{The anomalous average}

\begin{figure}  
\center
\includegraphics{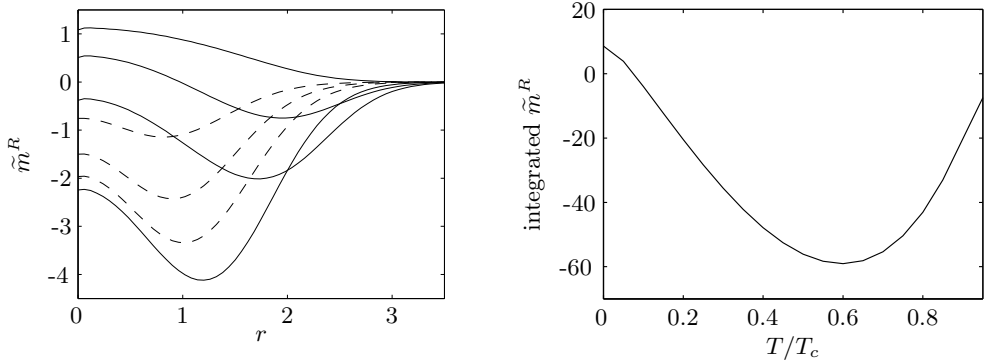}
\caption[Renormalized anomalous average at various temperatures and integrated anomalous
average as a function of temperature]{Renormalized anomalous average. \textbf{Left:}
$\mav^R$ at the temperatures $T/T_c$: 0, 0.15, 0.3, 0.75 (solid lines from top to
bottom), 0.85, 0.9 and 0.975 (dashed lines from bottom to top). The renormalization makes
$\mav^R$ positive at low temperatures. \textbf{Right:} Spatially integrated anomalous
average as a function of the non-interacting gas critical temperature. Note that
interactions decrease the critical temperature slightly \protect\cite{Cluso2}.}\label{fig:mtilde}
\end{figure}

In the left panel of Figure \ref{fig:mtilde} the renormalized anomalous average
$\mav^R$ is plotted at various temperatures. Often \mav\ is renormalized by
simply dropping the `1' in the term $2\fB +1$ in \eqref{mf_mt-uv}, which is
responsible for its divergence. Then the anomalous average is negative at all
times. At zero temperature, however, this implies $\mav \equiv 0$, which makes
the gapless HFB result for the many-body T-matrix useless. Utilizing the
semi-classical approximation to perform a more careful renormalisation, the
anomalous average becomes positive at low temperatures.

As the temperature is increased towards the critical temperature, the
condensate becomes highly depleted and the anomalous average vanishes. This is
shown in the right panel of Figure \ref{fig:mtilde}.

%% file: res_0K.tex
\subsection{Interaction strength at zero temperature}

To avoid confusion, we again point out that we refer to the many-body T-matrix
of section \ref{sec_tmb} as the \emph{quasi-2D coupling parameter}.  The
many-body T-matrix obtained numerically from the G2 theory is referred to as
the \emph{gapless HFB many-body T-matrix}.

\begin{figure}  
\center
\includegraphics{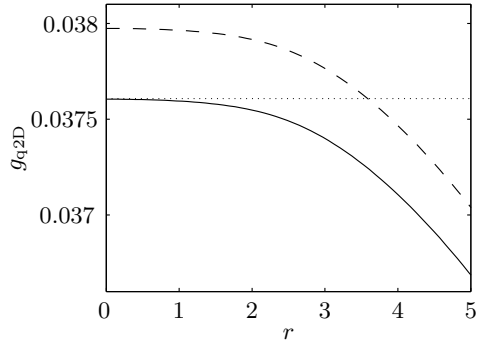}
\caption[Many-body T-matrix at zero temperature. First order approximation,
self-consistent solution and homogeneous limit of the quasi-2D coupling
parameter]{Quasi-2D coupling parameter at zero temperature. The full line
corresponds to the spatially dependent, self-consistent solution of
\eqref{cp:g2d_spatial}, the dashed line is the approximate solution
\eqref{cp:g2d_spatial_approx}. The dotted line is the homogeneous limit
\eqref{nm_g2D_TMB}. The dimensional parameter is $l_z/a_\mrm{3D} \approx 270$.
Note that the condensate density has dropped to below 1\% of its peak value at a
radius of 3.5 $l_\bot$ \protect\cite{Cluso2}.}\label{fig:gq2D}
\end{figure}

\begin{figure}  
\center
\includegraphics{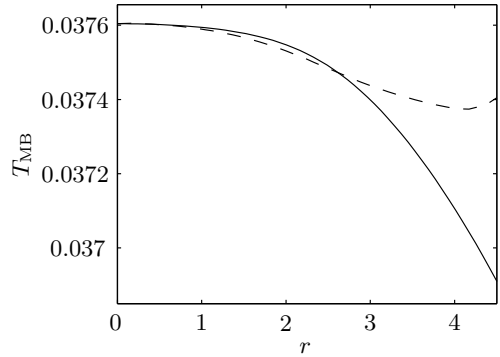}
\caption{Comparison of the quasi-2D coupling parameter (solid line) and the
gapless HFB many-body T-matrix (dashed line) at zero temperature. The gapless
HFB result has been calculated with $g$ chosen to achieve agreement in the trap
centre. The parameter of confinement is $l_z/a_\mrm{3D} \approx
270$.}\label{fig:TMB_cp_0K}
\end{figure}

Figure \ref{fig:gq2D} shows the quasi-2D coupling parameter for conditions
corresponding to the recent MIT experiment \cite{Goerlitz2001_2D}, where
$l_z/a_\mrm{3D} \approx 270$. In Figure \ref{fig:TMB_cp_0K} we compare the
self-consistent result from the previous figure with the result from gapless
HFB. The coupling strength $g$ in \eqref{mf_VconTMB} is chosen so that both
versions of the many-body T-matrix agree in the trap centre. This leads to very
good agreement between the two approaches in the region where interactions take
place. At the edge of the condensate the gapless HFB many-body T-matrix returns
to the value of $g$, which would be the two-body T-matrix in the
three-dimensional case. The quasi-2D coupling parameter approaches zero in the
case that $n_c=0$, however it remains finite in the numerical calculation due
to the weak logarithmic dependence on the condensate density.

\begin{figure}  
\center \hspace{8pt}
\includegraphics{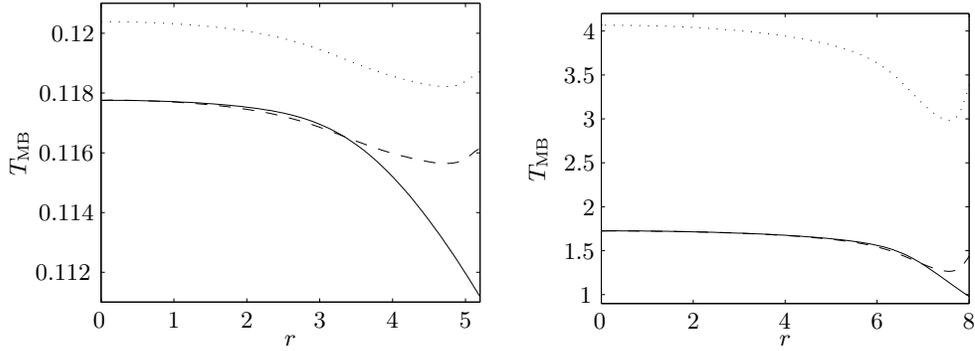}
\caption{Comparison of the off-shell two-body many-body T-matrix (solid line)
and the gapless HFB many-body T-matrix at zero temperature for different axial
trapping frequencies. The dotted line corresponds to the choice of $U_0 =
g'_\mrm{3D}$, the dashed line to $U_0 = g_\mrm{q2D}^\mrm{hom}$. With
increasing trapping frequency, the condensate approaches the purely
two-dimensional regime. In comparison to Figure \ref{fig:TMB_cp_0K}, the ratio
$l_z/a_\mrm{3D}$ is decreased to 84 (left) and 3 (right). This corresponds to
trapping frequencies of $10^1$ and $10^4$ times the original frequency of
$\omega_z = 2\pi\,790\,$Hz. The ordinate in all plots has been scaled to match
the extension of the condensate, \ie the radius of the condensate increases
with increasing $\omega_z$.}\label{fig:TMB_cp_frequencies}
\end{figure}

Figure \ref{fig:TMB_cp_frequencies} is similar to Figure \ref{fig:TMB_cp_0K},
but with the axial trapping frequency increased by a factor of $10^4$,
corresponding to an axial length scale $l_z/a_\mrm{3D}\approx 3$, which pushes
the condensate close to the genuine 2D limit.  The agreement between the
quasi-2D coupling parameter and the gapless HFB many-body T-matrix improves the
more `two-dimensional' the system becomes. Also shown is the gapless HFB
many-body T-matrix where we have used $g=g_\mrm{q3D}$ (dotted line). The
smaller the parameter $l_z/a_\mrm{3D}$, the larger the gap between the results
calculated using the quasi-2D and quasi-3D coupling parameters, indicating that
the quasi-3D parameter, which considers scattering to take place in three
dimensions, looses validity in this regime.

%% file: res_finT.tex
\subsection{Results at finite temperatures}

\begin{figure}  
\center
\includegraphics{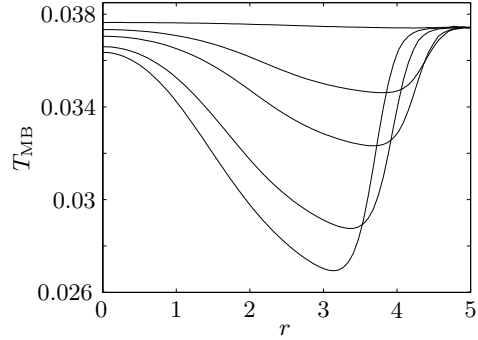}
\caption{Gapless HFB many-body T-matrix at 0, 0.3, 0.5, 0.75 and 0.85 $T/T_c$
from top to bottom.}\label{fig:tmb_finT}
\end{figure}

Finite temperature results for the many-body T-matrix, calculated within the gapless HFB
formalism, are depicted in Figure \ref{fig:tmb_finT}. The parameter $g$ in \eqref{mf_VconTMB}
does not depend on temperature, but only on the trap geometry, since finite temperature
contributions are naturally incorporated through the anomalous average and the condensate
density. Thus, we can determine $g$ at zero temperature from comparison with the off-shell
two-body T-matrix and use this parameter in the finite temperature calculations. This method
relies on the agreement of the quasi-2D coupling parameter and the gapless HFB many-body
T-matrix at zero temperature. In the previous section we have shown that these do agree in the
quasi-2D limit in the spatial regime where condensate interactions take place. Comparable
results for the gapless HFB many-body T-matrix have also been obtained for the three
dimensional case \cite{Hutchinson2000a}.

As can be seen in Figure \ref{fig:tmb_finT}, with increasing temperature and non-condensate
fraction many-body effects cause the interaction strength to drop down to about 70\% of its
maximal value within the extent of the condensate.

\begin{figure}  
\center
\includegraphics{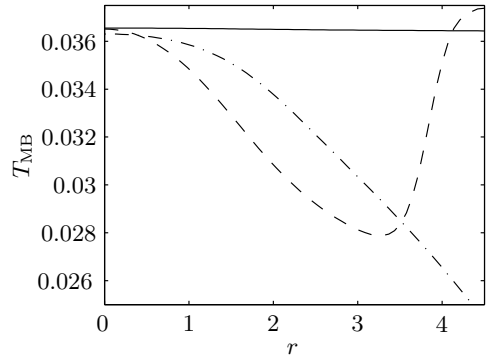}
\caption{Many-body T-matrix at 0.85$\,T_c$: The solid {\bf line} is the quasi-2D coupling
parameter from Figure \ref{fig:TMB_cp_0K}. The dashed line is the gapless HFB
many-body T-matrix from Figure \ref{fig:tmb_finT} with $g$ determined at zero
temperature as described in the text. The chained line corresponds to the
solution of the finite temperature extension \eqref{cp:g2d_finT_tosi} within
the local density approximation.}\label{fig:tmb_0.85}
\end{figure}

The approximation of the many-body T-matrix via the off-shell two-body T-matrix
in the zero-temperature limit cannot represent the behaviour of the gapless HFB
result, even if the population term in \eqref{cp:g2d_finT_tosi} is included. In
Figure \ref{fig:tmb_0.85} we explicitly show the discrepancy of the zero
temperature quasi-2D coupling parameter (solid line) and the finite temperature
gapless HFB many-body T-matrix (dashed line). If, however, we use the local
density approximation \eqref{finT_lda} to replace the chemical potential by the
density term from the GPE, we get much better agreement even at temperatures as
high as $0.85~T_c$ (chained line).

Still, the result based on the approach in \cite{Tosi2004a} seems to
underestimate finite-temperature many-body effects. We pointed out in Section
\ref{sec_finT_g2d} how problems occur in two dimensions when deriving the
T-matrix from the homogeneous case. The major contribution to scattering
effects comes from the highly occupied low-energy states. The
finite-temperature effects on these low-lying states are not properly accounted
for in the derivation of \eqref{cp:g2d_finT_tosi} from the homogeneous gas. Due
to the large population the Bose-enhancement amplifies the effect of these
states, and this is the reason for the discrepancy between the
two approaches.

%% file: conclusion.tex
\section{Discussion}\label{sec_conclusion}

The determination of the many-body T-matrix is important in lower dimensional
systems where scattering processes cannot be described merely by the two-body
T-matrix. The effects of the surrounding many-body medium on condensate
interactions must be taken into consideration at all temperatures.

In this paper we have presented results based on a gapless extension of the
full HFB theory, used previously in the three dimensional case
\cite{Hutchinson2000a}. The self-consistent HFB method allows the determination
of the many-body T-matrix for the whole temperature regime below the critical
temperature. Approaches complementary to the HFB method exist, and we compare
our results to a quasi-two-dimensional coupling parameter obtained from
evaluating the off-shell two-body T-matrix in two dimensions. Lee \etal have
shown that this corresponds to the many-body T-matrix if evaluated at a shifted
collision energy \cite{Lee2002a}. We find that the gapless HFB many-body
T-matrix agrees well with the results obtained from the off-shell two-body
T-matrix at zero temperature. Depending on the strength of the axial confining
potential, the agreement improves as the system changes from close to the
quasi-3D to the quasi-2D regime, as this is the relevant regime for the
calculation. The agreement between the two different approaches confirms the
validity of the semi-classical renormalization method we employ to remove the
ultra-violet divergence of the anomalous average.

On close examination of the finite temperature results, we see that the
many-body T-matrix develops a strong spatial dependence, decreasing in strength
towards the edge of the condensate. While finite temperature contributions to
many-body effects on scattering are neglected in the approximation used in
\cite{Lee2002a}, an extension to finite temperatures has recently been
presented for the homogeneous case by Rajagopal \etal \cite{Tosi2004a}. The
inclusion of a population factor due to the Bose distribution in
\eqref{cp:TMb=LippS} leads to a decrease in the homogeneous interaction
strength with increasing temperature, which we confirm. However, the previously
mentioned spatially dependent decrease is much more dominant than this shift.
By invoking a local density approach, we obtain much better agreement between
the gapless HFB many-body T-matrix and the finite temperature version of the
quasi-2D coupling parameter even at high temperatures, although the latter
underestimates finite-temperature many-body effects on scattering. The reason
for this we identify as an inappropriate treatment of the strongly contributing
low-energy states, originating in the derivation of the coupling parameter for
the homogeneous gas.  The phonon modes which have been neglected in the
treatment of Rajagopal \etal appear to be significant.  These modes are
included in the G2 numerical results.

The self-consistent gapless HFB calculation requires the choice of a temperature independent
two-body coupling strength to start with. Having shown that the gapless HFB and the off-shell
two-body T-matrix approach agree very well at zero temperature, we can determine this
parameter by matching the gapless HFB result to the quasi-2D coupling parameter. The parameter
depends only on the trap geometry and can, therefore, be used for the finite temperature
calculation where the result from \cite{Lee2002a} is no longer valid. Our approach therefore
renders a consistent technique which is easy to implement and valid for all temperatures below
the transition temperature.

In conclusion, we have shown that the many-body T-matrices determined from two very different
approaches, the gapless HFB theory with a semi-classically renormalized anomalous average and
the off-shell two-body T-matrix, agree very well at zero temperature and throughout a large
spatial regime at finite temperatures.

%% file: ack.tex
\ack

We would like to acknowledge the Marsden Fund of the Royal Society of New
Zealand and the Royal Society (London) for financial support. We would also
like to thank Sam Morgan for exceedingly useful discussions. 
